# Improvement of $H_2O_2$ electrogeneration using a Vulcan XC72 carbon-based electrocatalyst modified with Ce-doped $Nb_2O_5$

*Aline B. Trench [1], João Paulo C. Moura [1], Vanessa S. Antonin [1],*

*Caio Machado Fernandes [1], Liying Liu [2], Mauro C. Santos [1*]*

[1] *Laboratório de Eletroquímica e Materiais Nanoestruturados, Centro de Ciências Naturais e Humanas, Universidade Federal do ABC. Rua Santa Adélia 166, Bairro Bangu, 09210-170, Santo André - SP, Brasil.*

[2] *Centro Brasileiro de Pesquisas Físicas. Rua Dr. Xavier Sigaud 150, Bairro Botafogo, 22290-180, Rio de Janeiro – RJ, Brasil.*

*\*Corresponding author.* E-mail address: mauro.santos@ufabc.edu.br

# ABSTRACT

The use of the oxygen reduction reaction (ORR) for in-situ production of $H_2O_2$ is an attractive alternative to replace the methods based on anthraquinone oxidation. This study investigates the modification of Vulcan XC72 carbon with Ce-doped $Nb_2O_5$ in different molar proportions and its application as electrocatalysts in the ORR. One performed the characterization of the electrocatalysts using X-ray diffraction, Raman spectroscopy, scanning electron microscopy, transmission electron microscopy, contact angle measurements, and X-ray photoelectron spectroscopy. Subsequently, the electrocatalysts were analyzed for the ORR and the $Nb_2O_5$ doped with 0.5% Ce showing the highest electrocatalytic response. This electrocatalyst was also employed as a gas diffusion electrode and exhibited more significant $H_2O_2$ production at all potentials than the Vulcan XC72 carbon modified solely with $Nb_2O_5$. At the applied potentials of -1.3 V and -1.9 V, it produced 105% and 86% more $H_2O_2$, respectively, than the Vulcan XC72 carbon modified only with $Nb_2O_5$. These results can be attributed to the doping of $Nb_2O_5$ with 0.5% Ce, which induces local distortions in the crystal lattice of $Nb_2O_5$ due to the difference in ionic radius between $Nb^{5+}$ and $Ce^{3+}$, which combined with increased hydrophilicity and wetting properties, may have facilitated electron transfer and $O_2$ transport, favoring the ORR.

# 1. Introduction

Hydrogen peroxide ($H_2O_2$) is a versatile and environmentally friendly oxidizing agent due to its ability to decompose only into $H_2O$ and $O_2$. Due to its qualities, $H_2O_2$ has a domestic and industrial use. In particular, its usage in industry is for paper and cellulose manufacturing and disinfection processes [1, 2]. Currently, $H_2O_2$ is produced, to a large extent, by the anthraquinone oxidation process. This method is not considered safe and green, and its replacement is essential [3, 4].

The electrogeneration of hydrogen peroxide in situ using the oxygen reduction reaction (ORR) via two electrons is an alternative and environmentally friendly method for producing $H_2O_2$ (Eq. 1) [2, 5].

$$O_2 + 2e^- + 2H^+ \rightarrow H_2O_2 \qquad (1)$$

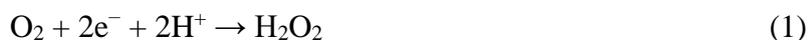

Materials such as platinum (Pt) [6] and palladium (Pd) [7] are considered active for ORR via 4 electrons, where there is no formation of $H_2O_2$, as shown in Eq. 2.

$$O_2 + 4H^+ + 4e^- \rightarrow 2H_2O \qquad (2)$$

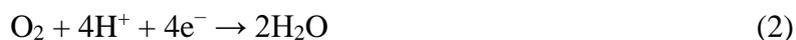

Materials such as gold (Au) and mercury (Hg) are more selective in reducing $O_2$ to $H_2O_2$ via the 2-electron reaction pathway [8]. In this aspect, researchers studied the coupling of materials such as Pt and Pd (active for ORR) with materials such as Au and Hg (selective for $H_2O_2$) [9, 10]. However, using these materials provokes high costs and low availability [1, 11].

In the search for alternative materials, carbonaceous materials gained prominence due to their low cost, excellent availability on the market, and selectivity for the production of $H_2O_2$. Its selectivity for the 2-electron mechanism is due to the presence of oxygenated functional groups that serve as active sites for the adsorption of the $O_2$ molecule [12-15]. Only by inserting very small amounts of metal oxides into carbonaceous materials can the electrocatalytic activity for generating H2O2 from ORR improve [11, 16-20].

This improvement in $H_2O_2$ electrogeneration has already been reported through the use of graphene modified with $ZrO_2$ [21], carbon black modified with $ZrO_2$ [22], $V_2O_5$ [23], $CeO_2$ [18], $MnO_2$ [24, 25], $WO_3$ [16], $NaNbO_3$ microcubes decorated with $CeO_2$ [26], among others. Metal oxides, when inserted into carbon matrices, can increase the active area of electrocatalysts and their hydrophilicity, facilitating the transport and adsorption of $O_2$. Furthermore, the synergistic effect between the materials can enhance their electrocatalytic activity.

Niobium oxide ($Nb_2O_5$) can improve the efficiency of carbonaceous materials, such as Vulcan XC72 carbon and reduced graphene oxide, as catalysts for the electrogeneration of $H_2O_2$ [27, 28]. Trench *et al*. [28] reported that carbon modified with $Nb_2O_5$ produced about 48 % more $H_2O_2$ than unmodified carbon. This result is due to the more hydrophilic character of electrocatalysts changed with $Nb_2O_5$ and the presence of a greater quantity of oxygenated species. Given this, it is clear that the insertion of metal oxides into carbonaceous materials is a promising alternative in the search for efficient options for producing $H_2O_2$.

A little-explored approach for $H_2O_2$ electrogeneration is doping a metallic oxide, which consists of intentionally introducing impurities, such as anions or cations, into a semiconductor to modify and improve its properties [29-31]. Cerium (Ce), one of the rare earth elements, stands out as a cationic dopant. This element produces strong magnetism and electrical properties ideal for different applications [32]. Furthermore, rare earths promote active sites and influence the electronic distribution of materials, improving the catalytic properties of materials [33, 34].

In this work, we propose a new electrocatalyst based on Vulcan XC72 carbon modified with Ce-doped $Nb_2O_5$ in different molar quantities (0.5, 1, and 2%) to electrogeneration $H_2O_2$ from ORR. For the first time, the influence of the doping of a metallic oxide on $H_2O_2$ electrogeneration is explored. Furthermore, this new electrocatalyst will also be applied in GDE for $H_2O_2$ electrogeneration, where its energy consumption, current efficiency and the amount of $H_2O_2$ generated will be assessed. The investigation is further solidified by analyzing the presence of oxygenated species, hydrophilicity, and the degree of disorder exhibited by this new electrocatalyst to understand how $Nb_2O_5$ doping affects the carbon matrix for $H_2O_2$ electrogeneration.

## 2. Material and methods

2.1. Preparation of the electrocatalysts

The synthesis of Ce-doped $Nb_2O_5$ nanorods followed the peroxide oxidation method [35], with crystallization obtained under hydrothermal conditions. In this procedure, determined amounts of $(NH_4[NbO(C_2O_4)_2(H_2O)]\cdot(H_2O)n$ (Sigma Aldrich) were introduced into 20 mL of distilled water with vigorous magnetic stirring at room temperature. Subsequently, $Ce(NO_3)_3 \cdot 6\ H_2O$ (Sigma Aldrich) was added to this solution in 0, 0.005, 0.01 and 0.02 moles and stirred for 10 minutes (min). Finally, the $H_2O_2$ (Sigma Aldrich) was then incorporated into the solution in the molar proportion $H_2O_2$ :

($NH_4[NbO(C_2O_4)2(H_2O)]·(H_2O)_n$ of 10:1, and the mixture was stirred vigorously for another 10 min. Subsequently, the solution was transferred to a hydrothermal reactor and maintained at 140°C for 12 hours (h). The resulting material was then cooled, subjected to centrifugation, washed with distilled water and isopropyl alcohol, and dried at 80°C for 12 h. The samples were named according to the amount of dopant: pure $Nb_2O_5$, Nb/Ce 0.5%, Nb/Ce 1.0%, and Nb/Ce 2.0% for 0.000, 0.005, 0.01 and 0.02 moles, respectively. Using the impregnation method, Ce-doped $Nb_2O_5$ nanoparticles, the electrocataysts prepared, were deposited on a Vulcan XC-72 carbon support [36]. 0.5 g of Vulcan XC-72 carbon was dispersed in 30 mL of $H_2O$ through magnetic stirring for 15 min. Specific amounts of $Nb_2O_5$ nanoparticles and Ce-doped $Nb_2O_5$ nanoparticles were measured to create electrocatalysts with 1% (by mass) of these nanoparticles and added to Vulcan XC-72 carbon dispersions and magnetically stirred for 4 h. Subsequently, the electrocatalysts were dried at 90°C in an oven. The resulting electrocatalysts were named as follows: VC/Nb for doping with 0.000 mol Ce, VC/Nb/Ce 0.5% for doping 0.005 mol Ce, VC/Nb/Ce 1.0 % for doping 0.01 mol, and VC/Nb/Ce 2.0 % for 0.02 mol doping. Additionally, the Vulcan XC 72 carbon-only electrocatalyst was denoted as VC XC72.

2.2. Characterization of electrocatalysts

X-ray diffraction (XRD) analyzes were carried out using two different equipment. The $Nb_2O_5$ samples and Ce-doped $Nb_2O_5$ samples were analyzed using a D8 Focus diffractometer manufactured by Bruker AXS. A Kα copper X-ray source (wavelength λ = 1.54 Å) was used, with the instrument operating in continuous scanning mode at a rate of 2 degrees per min, covering the range from 20 to 80 degrees (2θ). For carbon impregnated samples (VC XC72, VC/Nb/Ce 0.5%, VC/Nb/Ce 1.0% and VC/Nb/Ce 2.0%) it was used a Panalytical model X'Pert Pro-PW3042/10 diffractometer equipped with Cu Kα radiation (λ = 0.1540 nm) at 40 kV and 40 mA. A solid-state X-Celerator detector was employed. Scans were conducted over a 2θ range of 10 to 90 °C at a scanning rate of 0.025 ° s$^{-1}$.

Raman scattering analysis was conducted utilizing a Horiba-Jobin-Yvon model T64000 spectrometer, coupled with a 532 nm laser source, with an exposure duration of 40 seconds.

The morphological analysis was carried out using scanning electron microscopy (SEM), using a FESEM JEOL microscope model JSM-7401F. Detailed microstructural analyses were performed by using transmission electron microscopy (TEM) with JEOL JEM 2100 for $Nb_2O_5$ sample and JEOL JEM 2100F for VC/Nb/Ce 2.0% sample. Both TEMs operate at 200 kV. The energy-dispersive X-ray spectroscopy (EDS) was used to generate elementary maps in scanning transmission electron microscopy (STEM) mode.

To evaluate the hydrophilicity of the materials, contact angle measurements were carried out using a goniometer (GBX Digidrop). To carry out this analysis, 2 mg/mL of the material was prepared in distilled water subjecting them to sonication for 60 min in an ultrasonic bath. Subsequently, a 40 µL aliquot of each dispersion was deposited on a glassy carbon plate and dried to form a thin and uniform film. Then, one gently applied 10 mL of distilled water to the film surface to determine the contact angle. These measurements were carried out in triplicate using the Windrop software.

X-ray photoelectron spectroscopy (XPS) was performed using a Scienta Omicron ESCA + spectrometer (Germany), with monochromatic Al Kα X-rays (energy 1486.7 eV) as the radiation source. The inelastic background in the high-resolution center-level C 1s spectra was subtracted using the Shirley method. Spectral fitting was performed without imposing restrictions using multiple Voigt profiles, and CasaXPS software was used for data analysis.

## 2.3. Oxygen reduction reaction study

Electrochemical measurements were conducted using an Autolab PGSTAT-302N potentiostat/galvanostat, operated with the NOVA software, and a rotating ring disk electrode system (RRDE). To assess the oxygen reduction reaction (ORR), a cell with a 125 mL capacity was utilized. This setup included a Pt counter electrode with a surface area of 2 cm², an Hg/HgO reference electrode with an internal solution of 5 mol L$^{-1}$ NaOH, and the RRDE itself, serving as the working electrode.

The RRDE featured a disk made of vitreous carbon with an area of 0.2475 cm² and a gold (Au) ring with an area of 0.1866 cm². The experimental collection factor (N) was 0.28, while the theoretical collection factor was 0.22. The experiments were carried out in a 100 mL solution of 1 mol L$^{-1}$ NaOH (Synth) as the supporting electrolyte.

The VC XC72 carbon-based electrocatalysts modified with Ce-doped $Nb_2O_5$ and the VC XC72 electrocatalyst were applied to the RRDE disk as dispersions, produced at a ratio of 1 mg mL$^{-1}$ (material/distilled water) by employing high-end ultrasound for 1 min. Subsequently, 20 µL aliquots of the mixture were pipetted onto the vitreous carbon substrate, and the water was dried under a constant flow of nitrogen gas. Following this, 20 µL of Nafion solution (1:100 v/v, Nafion: distilled water, Sigma Aldrich) was deposited over the dried film and dried again.

For all electrochemical analyses, the electrolyte was first saturated with oxygen gas ($O_2$, White Martins) for 30 min, with the $O_2$ flow being maintained during the measurements above the solution. The measurements were conducted at a sweep rate of 5 mV s$^{-1}$ at room temperature. The ring electrode was polarized at 0.3 V (vs Hg/HgO) to ensure the oxidation of all $HO_2^-$ ions reaching it, while an anode sweep ranging from 0.1 V to -0.6 V (vs Hg/HgO) was applied to the disk electrode. The rotation rates, corresponding to 100, 400, 625, 900, and 1600 revolutions per min (rpm), were controlled by a CTV10 speed control unit [25, 26].

2.4. $H_2O_2$ electrogeneration

The gas diffusion electrode (GDE) was prepared using the electrocatalyst that exhibited the most promising electrocatalytic response for ORR. The GDE was constructed with a diameter of 2 cm and a thickness of 0.4 cm, providing an exposed surface area of 3 cm² between two porous steel plates. To create the GDE, 6 grams of the respective electrocatalyst was dispersed in 100 mL of $H_2O$ for 30 min.

Then, a poly(tetrafluoroethylene) (PTFE) solution, composed of 60 % PTFE by mass (Sigma Aldrich), was mixed with 20 mL of distilled water in a ratio of 4:1 (material: PTFE). This PTFE solution was gently introduced into the electrocatalyst dispersion under continuous stirring and left under stirring for another 40 min. The electrocatalyst mixture was vacuum-filtered and dried until its moisture content varied between 5% and 10%.

The electrocatalyst was then compressed using two steel plates at 290°C for 2 h, facilitated by a hydraulic heat press (SOLAB SL-10/15-E). An Autolab PGSTAT-302N potentiostat/galvanostat was used to regulate the potential applied during the electrogeneration of $H_2O_2$. Additionally, a multimeter was connected to the cathode and anode to measure the cell potential.

The electrochemical experiments were conducted using an electrochemical cell containing 350 mL of an electrolyte support solution composed of 0.1 mol L$^{-1}$ of $H_2SO_4$ (Synth) and 0.1 mol L$^{-1}$ of $K_2SO_4$ (Synth). The cell was equipped with a coil connected to a thermostatic bath to maintain the cell temperature at 20 °C. The configuration included an auxiliary Pt electrode (anode) with an area of 7.5 cm², an Ag/AgCl reference electrode (saturated KCl) and the GDE serving as the working electrode (cathode). The distance between the cathode and anode was set at 1 cm.

In experiments involving the electrogeneration of $H_2O_2$ through the GDE, various potentials (-0.7, -1.3, -1.9, and -2.5 V vs. Ag/AgCl) were applied for 2 h. Before using the potential, pressurized $O_2$ was purged at 0.2 bar directly in the GDE for 30 min to ensure that the electrolyte was saturated with $O_2$ during analysis. The progress of $H_2O_2$ electrogeneration was monitored by withdrawing 0.5 mL aliquots of the electrolyte at specific time intervals (0, 5, 10, 15, 20, 25, 30, 40, 50, 60, 75, 90, 105 and 120 min). These aliquots were then added to 4 mL of a solution of ammonium molybdate $((NH_4)_6Mo_7O_{24})$ 2.4 mmol $L^{-1}$ in $H_2SO_4$ 0.5 mol $L^{-1}$. The resulting samples were analyzed using a UV-Vis spectrophotometer (VARIAN 50 Scan), with absorbance at a wavelength of 350 nm, which was used to calculate the $H_2O_2$ concentration. The concentration of $H_2O_2$ was quantified in milligrams per liter (mg $L^{-1}$) with a correlation coefficient of 0.99.

## 3. Results and discussion

Fig. 1 shows a comparison of XRD patterns for various Nb/Ce samples with different mol percentages of Ce (0.0, 0.5, 1, and 2%). The diffraction peaks observed at $2\theta = 22.7°$, $35.0°$, $46.5°$, and $55.4°$ correspond to crystallographic planes (001), (101), (002), and (102) corresponding with the orthorhombic phase of $Nb_2O_5$ as defined by Joint Committee on Powder Diffraction Standards (JCPDS) with reference number 28-0317 [35, 37]. The peak observed at $2\theta \approx 27°$ was attributed to the presence of hydrated niobium oxide associated with intermediate crystallization conditions for $HNb_3O_8$ [37].

Notably, the XRD patterns for all doped samples maintained the same pattern as the pure $Nb_2O_5$ sample, suggesting that the introduction of Ce through doping did not alter the crystalline structure of $Nb_2O_5$. Furthermore, the absence of any peaks associated with the $CeO_2$ phase confirms the purity of the samples even after doping.

Comparing the Nb/Ce 0.5% sample with the pure $Nb_2O_5$ sample, it is evident that the intensity of the peaks increased with the insertion of 0.5 mol % Ce. When comparing the pure $Nb_2O_5$ sample with the Nb/Ce 1.0% and Nb/Ce 2.0% samples, the peak intensities decrease as the doping increases from 1.0 to 2.0%. This behavior suggests that the materials doped with 0.5% Ce have more excellent crystallinity and that doping with 1.0 and 2.0% Ce resulted in lower crystallinity [35, 38]. Furthermore, Ce doping caused a local distortion in the $Nb_2O_5$ crystal lattice as a result of the difference between the ionic radius of $Nb^{5+}$ (64 pm) and $Ce^{3+}$ (101 pm), inducing local distortions and defects in the $Nb_2O_5$ crystal lattice [38].

Fig. 1(b) shows the XRD pattern of VC XC72 used as a support for the fabrication of electrocatalysts based on Ce-doped $Nb_2O_5$ nanoparticles. It can be noted that the VC XC72 presents characteristic diffraction peaks at approximately $2\theta \approx 25°$ and $44°$[18]. Furthermore, Fig. 1(b) presents the XRD patterns of the VC/Nb/Ce 0.5%, VC/Nb/Ce 1.0%, and VC/Nb/Ce 2.0% electrocatalysts produced by the impregnation method. The electrocatalysts present the same diffraction peaks as VC XC72, which is the support material and is present in the majority of these electrocatalysts. Due to the low amount of $Nb_2O_5$ doped with Ce (1% by mass) and the detection limits of the equipment, the peaks related to Ce-doped $Nb_2O_5$ were not observed. However, the presence of $Nb_2O_5$ doped with Ce in the electrocatalysts produced was confirmed by other characterization techniques that will be presented throughout this discussion.

The Raman spectra of VC XC72, VC/Nb, and VC/Nb/Ce electrocatalysts were deconvoluted into four components and are shown in Fig. 2. The D1 band is associated with the disordered graphitic network found at the edges of graphene layers, and the G band represents the vibrational modes of the $sp^2$ sites within an ideal graphitic network. In the spectral region between bands D1 and G, around 1500 $cm^{-1}$, another band, called D3, corresponds to amorphous carbon. A band can be observed at approximately 1250 $cm^{-1}$, called the D4 band, which is attributed to $sp^2$-$sp^3$ bonds or stretching vibrations involving C = C and C − C bonds. The D3 and D4 bands are typically found in highly defective carbon materials [39, 40].

The ratio of the area of the D1 band and the area of the G band ($A_{D1}/A_G$) is used to evaluate the degree of disorder of the carbon networks. The $A_{D1}/A_G$ ratios were calculated and presented values of 0.88 and 1.09 for the VC XC72 and VC/Nb, respectively. The samples VC/Nb/Ce 0.5%, VC/Nb/Ce 1.0%, and VC/Nb/Ce 2.0% presented $A_{D1}/A_G$ ratios values of 1.12, 1.10, and 0.91, respectively [40].

It is observed that the ratio value increases from the VC XC72 sample to the VC/Nb sample, indicating that the insertion of $Nb_2O_5$ to the VC XC72 increased the degree of disorder. The VC/Nb/Ce electrocatalysts presented $A_{D1}/A_G$ ratio values higher than those of VC XC72, indicating that the degree of disorder of these samples also increased in relation to pure VC XC72. However, when we analyzed the VC/Nb/Ce 0.5% sample, it presented a higher $A_{D1}/A_G$ ratio compared to all other electrocatalysts presented in this study. In this way, the degree of disorder in this sample proved to be even more significant, indicating that the insertion of $Nb_2O_5$ nanoparticles doped with 0.5% Ce can generate a high degree of disorder in the carbon network. This result may suggest that the Ce-doped $Nb_2O_5$ electrocatalyst has a higher defect density compared to other electrocatalysts, which can improve electron transfer, positively affecting the ORR activity [25, 41].

The SEM micrographs of the $Nb_2O_5$ and Ce-doped $Nb_2O_5$ samples are shown in Fig. 3. It can be noted that the $Nb_2O_5$ sample is composed of agglomerated particles, as already reported by Trench *et al* [28]. The Ce-doped samples do not show significant morphological change, and all doped samples maintained the same morphology as the pure $Nb_2O_5$ sample, characterized by non-uniform particle agglomeration.

Fig. 4 shows TEM images for a pure $Nb_2O_5$ sample and VC/Nb/Ce 2.0%. The low magnification image in Fig. 4(a) shows similar agglomeration of particles as observed by SEM images. The higher magnification TEM image (Fig.4(b)) clearly demonstrates that the particles have a nanorod shape [28, 42]. Fig. 4(c) presents TEM image of VC/Nb/Ce 2.0%. The presence of spherical particles and particle clusters are noted, which can be attributed to the presence of VC XC72 and $Nb_2O_5$. TEM images at higher magnifications (Fig 4(c, d)) show the presence of nanorod-shaped particles, which might be attributed to the presence of $Nb_2O_5$.

In order to confirm the presence of $Nb_2O_5$ and Ce in the manufactured electrocatalysts, an elemental composition analysis by EDS mapping of the VC/Nb/Ce 2.0% sample was carried out and the results are shown in Fig. 5(a-e). The spherical particles are mainly composed by C, which are the VC XC72 matrix, as mentioned previously. The cluster of nanorods contain Nb, O, and Ce, with a homogeneous distribution of these elements. These results agree with those previously observed, confirming the incorporation of Ce into the structure of $Nb_2O_5$ as dopant. Furthermore, the impregnation method used to produce the VC/Nb/Ce electrocatalysts also proved to be efficient.

The contact angle measurements represented in Table 1 were employed to investigate the surface hydrophilicity of the electrocatalyst. It can be noted that the VC XC72 presented a contact angle of 39.35 degrees after the insertion of $Nb_2O_5$ into the VC XC72. When analyzing the contact angles of the doped samples, it can be noted that the contact angle value decreased for doping with 0.5% Ce concerning the pure VC/Nb electrocatalyst. Subsequently, the contact angle values increased again. for electrocatalysts doped with 1.0 and 2.0% Ce.

In this way, it can be established from the results that the insertion of $Nb_2O_5$ into the VC XC72 leads to an increase in the hydrophilicity of the electrocatalyst. However, inserting $Nb_2O_5$ doped with 0.5% Ce produces even more hydrophilic electrocatalysts, directly affecting the electrocatalytic response for $H_2O_2$ electrogeneration. More hydrophilic electrocatalysts may have more significant amounts of oxygenated species and more excellent wettability, which can facilitate the transport and adsorption of $O_2$, favoring ORR [25, 26, 43].

**Table 1.** Average contact angles of the bare VC XC72, VC/Nb, and VC/Nb/Ce with different amounts of dopant.

| Electrocatalyst | Contact Angle (°) |
|---|---|
| VC XC72 | 39.35 ± 0.51 |
| VC/Nb | 34.13 ± 1.70 |
| VC/Nb/Ce 0.5% | 32.41 ± 1.08 |
| VC/Nb/Ce 1.0% | 32.49 ± 1.13 |
| VC/Nb/Ce 2.0% | 62.30 ± 4.90 |

Fig. 6 (a) displays high-resolution XPS C 1s core-level spectra for VC/Nb/Ce 0.5 % sample. This spectrum was deconvolved into six components, each related to a phase present in carbon. The aromatic C-C phase (CC $sp^2$) is located at a binding energy of about 284 eV and the aliphatic C-H phase (CC $sp^3$) at approximately 285 eV. Different oxygenated groups such as C-O, C=O and O-C=O are located at 286.8 eV, 288 eV and 289 eV, respectively. The plasmonic transitions are located at 291 eV [44].

Fig. 6 (b) shows the concentration of oxygenated species reported by Trench *et al.*[28] for the VC XC72 and VC/Nb samples and those found for the VC/Nb/Ce 0.5% electrocatalyst, developed in this study. It can be noted that the concentration of oxygenated species increases from 11.49 at.% for the pure VC XC72 electrocatalyst to 15.66 at.% for the electrocatalyst modified with $Nb_2O_5$ [28]. When we analyzed the electrocatalyst changed with $Nb_2O_5$ doped with 0.5% Ce, we again observed an increase in oxygenated species, reaching 18.13%. This increased presence of oxygenated species in the $Nb_2O_5$ electrocatalyst doped with 0.5% Ce means that this proposed new electrocatalyst presents a more hydrophilic surface than other electrocatalysts. This result aligns with the contact angle measurements, indicating a higher degree of hydrophilicity for this electrocatalyst.

Fig. 6 (c-e) shows the deconvoluted O 1s spectra for the samples VC XC72, VC/Nb and VC/Nb/Ce 0.5%. It can be observed that the O 1s spectra are dominated by two components attributed to the C=O and O-C=O groups, located at approximately 532 eV and 533 eV, respectively [45, 46]. These groups were also identified in the C spectrum, indicating good agreement between the results found. Another component identified at approximately 534 eV is related to the presence of molecular water [46]. For the electrocatalysts modified with $Nb_2O_5$, an additional component of low binding energy (530 eV) related to the O-Nb bonds was attributed [47]. The VC XC72, VC/Nb, and VC/Nb/Ce 0.5% electrocatalysts presented a concentration of C=O and O-C=O groups of 83.3, 91.57 and 94.34 at.%, respectively. Furthermore, the O-Nb group presents a concentration of 2.98 and 1.15 at.% in the VC/Nb and VC/Nb/Ce 0.5% electrocatalysts, respectively. These results corroborate those discussed in the C spectrum and reinforce the presence of more oxygenated species in the 0.5% VC/Nb/Ce electrocatalyst.

Fig. 6(f) presents the deconvoluted spectra of Nb 3d for the VC/Nb and VC/Nb/Ce 0.5% electrocatalysts. The spectra were deconvoluted into two main peaks corresponding to Nb $3d_{5/2}$ and Nb $3d_{3/2}$ and with a spin-orbit separation of 2.7 eV corresponding to the $Nb^{5+}$ oxidation state [47, 48]. It can be noted that the Nb 3d peak in the VC/Nb/Ce 0.5% electrocatalyst showed a shift towards higher binding energy compared to the undoped electrocatalyst. This result highlights the effect of Ce doping on the structure of $Nb_2O_5$, indicating that the Ce dopant may be occupying the local structure of Nb, causing this shift to higher binding energies [49, 50].

The electrocatalytic performance of VC/Nb and VC/Nb/Ce electrocatalysts with different amounts of Ce doping was evaluated for the ORR in an alkaline medium (1 mol $L^{-1}$ NaOH). This evaluation was carried out using the Rotating Ring-Disk Electrode (RRDE) technique, focusing on generating steady-state polarization curves. The resulting data is presented in Fig. 7. The electrocatalytic activity of reference materials, specifically VC XC72 and Pt, was also evaluated under the 2-electron and 4-electron ORR mechanisms, respectively.

The curves depicted in Fig. 7 were constructed under a scan rate of 5 mV $s^{-1}$, with the RRDE set to rotate at 1600 rpm. The polarization curves of the VC/Nb and VC/Nb/Ce electrocatalysts showed disc currents close to those of VC XC72 and far from the current achieved with Pt. This observation indicates that the electrocatalysts presented in this study predominantly follow the 2-electron mechanism for ORR.

The VC/Nb and VC/Nb/Ce electrocatalysts produced more significant amounts of $H_2O_2$ than VC XC72 and Pt. Specifically, the VC/Nb/Ce 0.5% electrocatalyst exhibited a current of approximately 75 μA, while the VC XC72 was about 50 μA and VC/Nb was about 60 μA.

Furthermore, the polarization curve of VC/Nb/Ce 0.5% electrocatalysts exhibited a more positive initial potential for $H_2O_2$ electrogeneration compared to VC XC72. This suggests that applying this electrocatalyst can reduce energy consumption and, consequently, make the electrocatalyst proposed in this study more efficient compared to pure carbon.

The ORR data for all electrocatalysts were subjected to analysis using the Koutecky-Levich (K-L) Equation (Eq. 5). In this equation, 'i' represents the measured current, '$i_k$' and '$i_d$' denote the kinetic and diffusion-limited currents, respectively. 'k' stands for the rate constant for the ORR, 'F' corresponds to the Faraday constant (96,484 C mol$^{-1}$), 'A' represents the effective projected area covered by the catalyst, 'ω' is the rotation rate, '$C^0$' signifies the saturated oxygen concentration in the solution's bulk, '$D_{O2}$' is the oxygen diffusion coefficient, and 'ν' refers to the kinematic viscosity of the solution [36].

$$\frac{1}{i} = \frac{1}{i_k} + \frac{1}{i_d} = \frac{1}{nFAkC^0} + \frac{1}{\{0.62\, nFAD_{O_2}^{\frac{2}{3}} v^{-\frac{1}{6}} C^0 \omega^{\frac{1}{2}}\}} \qquad (3)$$

KL plots were generated by plotting 1/i against 1/ω$^{-1/2}$ using data collected from disc curves obtained at various rotational speeds (100, 400, 625, 900, and 1600 rpm). The graph in Fig. 8 shows that the VC/Nb and VC/Nb/Ce electrocatalysts with different doping amounts exhibit similar slopes to those of the VC XC72. However, they differ significantly from the slope of the Pt reference.

Considering that KL slope is inversely related to the number of electron transfers in the ORR and knowing that VC XC72 and Pt serve as reference materials for the 2- and 4-electron mechanisms, respectively, for the ORR, it can be inferred that the electrocatalysts developed closely align with the 2-electron ORR mechanism since we observed parallelism between the slopes of the VC/Nb/Ce and VC XC72 electrocatalysts.

The percentages of $H_2O_2$ and $H_2O$ production and the electron transfer number ($n_t$) were determined by analyzing the polarization curves obtained at 1600 rpm. This analysis was performed using Eq. 4 to 6 below, in which $i_r$ represents the ring current, $i_d$ represents the disc current and N represents the current collection efficiency of the ring made of gold (with N equal to 0,28) [36].

$$H_2O_2\% = \frac{200\ ir/N}{id\ +\ ir/N} \quad (4)$$

$$H_2O\ \% = 100 - H_2O_2\% \quad (5)$$

$$n_t = \frac{4id}{id\ +\ ir/N} \quad (6)$$

The calculation results for all electrocatalysts investigated are presented in Table 2. As expected, as it is a reference material for the 2-electron mechanism, VC XC72 exhibited an $n_t$ value close to 2. The proposed electrocatalysts VC/Nb/Ce and VC/Nb electrocatalyst also presented an $n_t$ value close to 2, indicating that these materials follow to the 2-electron mechanism, a conclusion that aligns with those previously seen in KL graphs. The Pt material presented an $n_t$ value close to 4, which was anticipated, as it served as a reference for the 4-electron mechanism in ORR.

Furthermore, it is noted that the VC/Nb and VC/Nb/Ce electrocatalysts showed greater amounts of $H_2O_2$ production compared to the pure VC XC72. In particular, VC/Nb/Ce 0.5% produced 84% $H_2O_2$, being the most prominent electrocatalyst. This result is in agreement with what was observed in the contact angle values, since this electrocatalyst presented a more excellent hydrophilic character, which is directly related to the ORR, since more hydrophilic electrocatalysts can provide more significant amounts of oxygenated species and More excellent wettability, facilitating transport and $O_2$ adsorption. This electrocatalyst also presented a higher concentration of oxygenated species compared to the VC XC72 and VC/Nb electrocatalysts in the XPS analyses, which positively affected its efficiency for RRO, making it more selective for $H_2O_2$ electrogeneration. It is also important to highlight that the electrogeneration of $H_2O_2$ is reduced as the amount of dopant increases. The ideal value was achieved with $Nb_2O_5$ doped with only 0.5% Ce.

**Table 2.** Number of electrons transferred ($n_t$) and conversion rate from $O_{2(g)}$ to $H_2O$ and $H_2O_2$ in ORR.

| Electrocatalyst | $n_t$ | $H_2O$ (%) | $H_2O_2$ (%) |
|---|---|---|---|
| VC XC72 | 2.6 | 29.7 | 70.3 |
| VC/Nb | 2.3 | 17.0 | 83.0 |
| VC/Nb/Ce 0.5% | 2.3 | 16.0 | 84.0 |
| VC/Nb/Ce 1.0% | 2.4 | 19.0 | 81.0 |
| VC/Nb/Ce 2.0% | 2.5 | 23.3 | 76.7 |
| Pt | 3.7 | 87.0 | 13.0 |

Fig. 9 illustrates the concentration of $H_2O_2$ electrogenerated in milligrams per liter (mg $L^{-1}$) throughout electrolysis at various potentials, using GDE manufactured with the VC/Nb/Ce 0.5 % electrocatalyst, which was the electrocatalyst that presented the highest efficiency, named VC/Nb /Ce0.5GDE. It can be noted that there was a production of approximately 763 mg $L^{-1}$ of $H_2O_2$ at the potential of -2.5 V. In a recent study, Trench *et al*.[28] reported output of 645 mg $L^{-1}$ at the same potential using Vulcan carbon GDE modified only with $Nb_2O_5$, named Nb1GDE.

Another factor evaluated in this study was energy consumption (EC, in kWhkg-1) and current efficiency (EC, in%), using Eqs 7 and 8, respectively [51]. For Eq. 7, I is the current measured during electrolysis (in A), Ecel is the cell potential (in V), t is the electrolysis time (in h), and m is the mass of the electrogenerated $H_2O_2$ (in kg ) [51].

For Eq.8, z is considered the number of electrons in the ORR intended for the electrogeneration of $H_2O_2$ (z = 2), F is the Faraday constant, $C_{H_2O_2}$ is the concentration of electrogenerated $H_2O_2$ (in g $L^{-1}$), V is the volume of solution (in L), $M_{H_2O_2}$ the molar mass of $H_2O_2$ (34.01 g $mol^{-1}$), I the current measured during electrolysis (in A), and t the electrolysis time (in s) [51].

$$EC = \frac{I\ E_{cel}\ t}{1000\ m} \quad (7)$$

$$CE = \frac{zFC_{H_2O_2}\ V}{M_{H_2O_2}\ It}\ 100 \quad (8)$$

Table 2 presents the results obtained for EC, EC, and the amount of $H_2O_2$ for VC/Nb /Ce0.5GDE after 120 min of electrolysis under application at different potentials. Furthermore, the table also shows the results obtained by Trench *et al*. [28] for Nb1GDE using the same experimental parameters to compare and analyze how the GDE modified with $Nb_2O_5$ doped with Ce behaves concerning the GDE changed with only $Nb_2O_5$. As can be seen, the GDE modified with $Nb_2O_5$ doped with Ce (VC/Nb /Ce0.5GDE) generated more significant amounts of $H_2O_2$ at all applied potentials than the GDE changed with only $Nb_2O_5$ (Nb1GDE). It is also worth mentioning that the VC/Nb /Ce0.5GDE was higher compared to the GDE of pure Vulcan XC72 carbon [28].

**Table 3.** Energy consumption (EC) and current efficiency (CE) were obtained for VC/Nb/Ce1GDE based on their application at different potentials.

| GDE | Applied potential (vs. Ag/AgCl) | $H_2O_2$ (mg L$^{-1}$) | EC (kWh kg$^{-1}$) | CE (%) | Reference |
|---|---|---|---|---|---|
| Nb1GDE | -2.5 V | 645.00 | 20.00 | 40.00 | [28] |
| VC/Nb/Ce0.5GDE | -2.5 V | 763.14 | 19.83 | 46.32 | This work |
| Nb1GDE | -1.9 V | 365.00 | 21.00 | 33.00 | [28] |
| VC/Nb/Ce0.5GDE | -1.9 V | 678.57 | 18.08 | 48.27 | This work |
| Nb1GDE | -1.3 V | 241.00 | 13.00 | 42.00 | [28] |
| VC/Nb/Ce0.5GDE | -1.3 V | 496.36 | 9.40 | 65.08 | This work |
| Nb1GDE | -0.7 V | 82.00 | 9.50 | 43.00 | [28] |
| VC/Nb/Ce0.5GDE | -0.7 V | 123.67 | 8.82 | 57.55 | This work |

Furthermore, for all applied potentials, VC/Nb /Ce0.5GDE presented lower EC and higher CE values than Nb1GDE. This shows that the new electrocatalyst proposed in this study, based on the doping of $Nb_2O_5$ with Ce, is very promising, producing more significant quantities of $H_2O_2$ with lower energy consumption and greater current efficiency compared to the pure $Nb_2O_5$ material.

It is worth mentioning that at the applied potential of -1.3 V and -1.9 V, VC/Nb /Ce0.5GDE produces around 105% and 86%, respectively, more $H_2O_2$ than Nb1GDE at the same applied potentials, highlighting the potential of the proposed electrocatalyst in this study.

Table 4 provides some studies reported in the literature using different GDE in experimental conditions similar to the one used in this article. It can be noted that the electrocatalyst proposed in this article presents superior results for the electrogeneration of $H_2O_2$, showing that the electrocatalyst based on $Nb_2O_5$ doped with 0.5 % of Ce is very promising for 2-electron RRO.

**Table 4.** $H_2O_2$ electrogeneration data using different GDE.

| GDE | Experimental parameters | $H_2O_2$ (mg L$^{-1}$) | Reference |
|---|---|---|---|
| 5% Co-Porphyrin/ PL6 | $K_2SO_4$ 0.1 mol L$^{-1}$, 90 min of electrolysis, at -1.75 V (vs Ag/Ag/Cl). | 340 | [52] |
| 1% WO$_3$/C supported on Vulcan XC 72R | 0.1 mol L$^{-1}$ $K_2SO_4$ and 0.1 mol L$^{-1}$ $H_2SO_4$ 120 min of electrolysis, at -1.3 V (vs Ag/Ag/Cl). | 585 | [44] |
| 1% α-MnO$_2$/ Vulcan XC72 | $H_2SO_4$ 0.1 mol L$^{-1}$ + $K_2SO_4$ 0.1 mol L$^{-1}$, 120 min of electrolysis, at -1.9 V (vs Ag/Ag/Cl). | 402 | [25] |
| 5% Ta$_2$O$_5$/PL6 | $K_2SO_4$ 0.1 mol L$^{-1}$, 120 min of electrolysis, at -1.0 V (vs Ag/Ag/Cl). | 27.9 | [45] |
| VC/Nb /Ce0.5GDE | $H_2SO_4$ 0.1 mol L-1 + $K_2SO_4$ 0.1 mol L$^{-1}$, 120 min of electrolysis, at -1.9 V (vs Ag/Ag/Cl). | 678.57 | This work |

In this sense, it is believed that doping $Nb_2O_5$ with 0.5% of Ce caused local distortions in the $Nb_2O_5$ crystal lattice due to the difference between the ionic radius of $Nb^{5+}$ (64 pm) and $Ce^{3+}$ (101 pm), inducing local distortions and defects in the $Nb_2O_5$ crystal lattice. This degree of disorder can be observed in Raman's analysis, where the VC/Nb/Ce 0.5% sample showed a greater degree of disorder in the carbon network compared to VC XC72 and VC/Nb. Furthermore, this sample presented the most excellent hydrophilic character and significant amounts of oxygenated species, observed in the contact angle and XPS analyses. Thus, the densities of defects generated in the 0.5 % Ce doped electrocatalyst combined with more excellent hydrophilicity and wettability and more significant quantities of oxygenated species may have facilitated the transfer of electrons, transport and adsorption of $O_2$, favoring the ORR and generating more substantial amounts of $H_2O_2$ [17, 25, 26, 51].

## 4. Conclusions

In the present work, Ce-doped $Nb_2O_5$ was successfully synthesized in different mol quantities (0, 0.5, 1.0, and 2.0%). XRD analyses showed that no secondary phase of $CeO_2$ was observed, indicating an effective doping of Ce in $Nb_2O_5$ for all samples. The doped materials were subsequently impregnated in Vulcan XC72 carbon, characterized by Raman spectroscopy, XPS, and contact angle, and tested for oxygen reduction reaction. All $Nb_2O_5$ electrocatalysts doped with Ce exhibited superior performance in the 2-electron ORR mechanism compared to Vulcan XC72 carbon without modification, highlighting the doping of $Nb_2O_5$ with 0.5% Ce. Based on the results observed for ORR, $H_2O_2$ electrogeneration experiments were conducted using a gas diffusion electrode with the most promising electrocatalyst for ORR (VC/Nb/Ce 0.5%). Our findings indicated that the application of VC/Nb/Ce0.5GDE led to a notable increase in $H_2O_2$ production at all potentials compared to the GDE of undoped $Nb_2O_5$. Furthermore, the energy consumption and current efficiency were lower and higher when compared to the $Nb_2O_5$ electrocatalyst without doping. This evidences that this new proposed electrocatalyst generates more significant quantities of $H_2O_2$ at a lower energy cost and greater current efficiency. It is also worth mentioning that at the applied potentials of -1.3 V and -1.9 V, VC/Nb /Ce0.5GDE produces around 105% and 86% more $H_2O_2$ than GDE of $Nb_2O_5$ without doping at the same applied potentials. These results can be attributed to the doping of $Nb_2O_5$ with 0.5% Ce, which causes local distortions in the $Nb_2O_5$ crystal lattice due to the difference between the ionic radius of $Nb^{5+}$ (64 pm) and $Ce^{3+}$ (101 pm), inducing local distortions and defects in the $Nb_2O_5$ crystal structure. This degree of disorder was observed in Raman's analysis, where the VC/Nb/Ce 0.5% sample showed a greater degree of disorder in the carbon network compared to the Vulcan XC72 carbon and Vulcan XC72 carbon modified with $Nb_2O_5$ electrocatalysts. Furthermore, VC/Nb/Ce 0.5% presented the most excellent hydrophilic character and more significant amounts of oxygenated species, as observed in the contact angle and the XPS analyses. Thus, the densities of defects generated in the electrocatalyst doped with 0.5% Ce combined with more excellent hydrophilicity and wettability and more significant amounts of oxygenated species may have facilitated the transfer of electrons, transport and adsorption of $O_2$, favoring the ORR and generating more significant amounts of $H_2O_2$.

**CRediT authorship contribution statement**

**Aline B. Trench,** Conceptualization**,** Investigation, Validation, Data curation, Writing – original draft, review & editing**, João Paulo C. Moura, Vanessa S. Antonin, Caio Machado Fernandes:** Investigation, Validation, Data curation, Writing – original draft, review & editing. **Mauro C. Santos:** Conceptualization, Writing – review & editing, Supervision.

**Acknowledgments**

The authors are grateful to the following Brazilian research financing institutions: São Paulo Research Foundation (FAPESP grants, #2021/14394-7, #2021/05364-7, and # 2022/10484-4) and the Coordination for the Improvement of Higher Education Personnel (CAPES, 88887.354751/2019-00), for the financial assistance provided in support of this work.

# References

[1] J.M. Campos-Martin, G. Blanco-Brieva, J.L. Fierro, Hydrogen peroxide synthesis: an outlook beyond the anthraquinone process, Angewandte Chemie, 45 (2006) 6962-6984.
[2] W. Zhou, X. Meng, J. Gao, A.N. Alshawabkeh, Hydrogen peroxide generation from O2 electroreduction for environmental remediation: A state-of-the-art review, Chemosphere, 225 (2019) 588-607.
[3] W. Zhou, L. Rajic, L. Chen, K. Kou, Y. Ding, X. Meng, Y. Wang, B. Mulaw, J. Gao, Y. Qin, A.N. Alshawabkeh, Activated carbon as effective cathode material in iron-free Electro-Fenton process: Integrated H(2)O(2) electrogeneration, activation, and pollutants adsorption, Electrochim Acta, 296 (2019) 317-326.
[4] S. Yang, A. Verdaguer-Casadevall, L. Arnarson, L. Silvioli, V. Čolić, R. Frydendal, J. Rossmeisl, I. Chorkendorff, I.E.L. Stephens, Toward the Decentralized Electrochemical Production of $H_2O_2$: A Focus on the Catalysis, ACS Catalysis, 8 (2018) 4064-4081.
[5] E. Yeager, Electrocatalysts for O2 reduction, Electrochim. Acta 29 29 (1984) 1527–1537.
[6] D. Wu, X. Shen, Y. Pan, L. Yao, Z. Peng, Platinum Alloy Catalysts for Oxygen Reduction Reaction: Advances, Challenges and Perspectives, ChemNanoMat, 6 (2019) 32-41.
[7] G.V. Fortunato, E. Pizzutilo, A.M. Mingers, O. Kasian, S. Cherevko, E.S.F. Cardoso, K.J.J. Mayrhofer, G. Maia, M. Ledendecker, Impact of Palladium Loading and Interparticle Distance on the Selectivity for the Oxygen Reduction Reaction toward Hydrogen Peroxide, The Journal of Physical Chemistry C, 122 (2018) 15878-15885.
[8] V. Viswanathan, H.A. Hansen, J. Rossmeisl, J.K. Norskov, Unifying the 2e(-) and 4e(-) Reduction of Oxygen on Metal Surfaces, J Phys Chem Lett, 3 (2012) 2948-2951.
[9] G.V. Fortunato, L.S. Bezerra, E.S.F. Cardoso, M.S. Kronka, A.J. Santos, A.S. Greco, J.L.R. Junior, M.R.V. Lanza, G. Maia, Using Palladium and Gold Palladium Nanoparticles Decorated with Molybdenum Oxide for Versatile Hydrogen Peroxide Electroproduction on Graphene Nanoribbons, ACS applied materials & interfaces, 14 (2022) 6777-6793.
[10] J.S. Jirkovsky, I. Panas, E. Ahlberg, M. Halasa, S. Romani, D.J. Schiffrin, Single atom hot-spots at Au-Pd nanoalloys for electrocatalytic $H_2O_2$ production, Journal of the American Chemical Society, 133 (2011) 19432-19441.
[11] M.S. Kronka, P.J.M. Cordeiro-Junior, L. Mira, A.J. dos Santos, G.V. Fortunato, M.R.V. Lanza, Sustainable microwave-assisted hydrothermal synthesis of carbon-supported $ZrO_2$ nanoparticles for $H_2O_2$ electrogeneration, Materials Chemistry and Physics, 267 (2021) 124575.
[12] P. Petsi, K. Plakas, Z. Frontistis, I. Sirés, A critical assessment of the effect of carbon-based cathode properties on the in situ electrogeneration of $H_2O_2$, Electrochimica Acta, 470 (2023) 143337.
[13] X. Hu, X. Zeng, Y. Liu, J. Lu, X. Zhang, Carbon-based materials for photo- and electrocatalytic synthesis of hydrogen peroxide, Nanoscale, 12 (2020) 16008-16027.
[14] J. An, Y. Feng, Q. Zhao, X. Wang, J. Liu, N. Li, Electrosynthesis of H(2)O(2) through a two-electron oxygen reduction reaction by carbon based catalysts: From mechanism, catalyst design to electrode fabrication, Environmental science and ecotechnology, 11 (2022) 100170.
[15] N. Li, C. Huang, X. Wang, Y. Feng, J. An, Electrosynthesis of hydrogen peroxide via two-electron oxygen reduction reaction: A critical review focus on

hydrophilicity/hydrophobicity of carbonaceous electrode, Chemical Engineering Journal, 450 (2022) 138246.

[16] M.L.O. Machado, E.C. Paz, V.S. Pinheiro, R.A.S. de Souza, A.M.P. Neto, I. Gaubeur, M.C. dos Santos, Use of WO2.72 Nanoparticles/Vulcan® XC72 GDE Electrocatalyst Combined with the Photoelectro-Fenton Process for the Degradation of 17α-Ethinylestradiol (EE2), Electrocatalysis, (2022).

[17] E. Paz, V. Pinheiro, L. Aveiro, F. Souza, M. Lanza, M. Santos, Hydrogen Peroxide Electrogeneration by Gas Diffusion Electrode Modified With Tungsten Oxide Nanoparticles for Degradation of Orange II and Sunset Yellow FCF Azo Dyes, Journal of the Brazilian Chemical Society, (2019).

[18] V.S. Pinheiro, E.C. Paz, L.R. Aveiro, L.S. Parreira, F.M. Souza, P.H.C. Camargo, M.C. Santos, Ceria high aspect ratio nanostructures supported on carbon for hydrogen peroxide electrogeneration, Electrochimica Acta, 259 (2018) 865-872.

[19] V.S. Antonin, M.H.M.T. Assumpção, J.C.M. Silva, L.S. Parreira, M.R.V. Lanza, M.C. Santos, Synthesis and characterization of nanostructured electrocatalysts based on nickel and tin for hydrogen peroxide electrogeneration, Electrochimica Acta, 109 (2013) 245-251.

[20] L.C. Trevelin, R.B. Valim, J.C. Lourenço, A. De Siervo, R.S. Rocha, M.R.V. Lanza, Using ZrNb and ZrMo oxide nanoparticles as catalytic activity boosters supported on Printex L6 carbon for H2O2 production, Advanced Powder Technology, 34 (2023) 104108.

[21] J.F. Carneiro, M.J. Paulo, M. Siaj, A.C. Tavares, M.R.V. Lanza, Zirconia on Reduced Graphene Oxide Sheets: Synergistic Catalyst with High Selectivity for H2O2 Electrogeneration, ChemElectroChem, 4 (2017) 508-513.

[22] M.S. Kronka, G.V. Fortunato, L. Mira, A.J. dos Santos, M.R.V. Lanza, Using Au NPs anchored on ZrO2/carbon black toward more efficient H2O2 electrogeneration in flow-by reactor for carbaryl removal in real wastewater, Chemical Engineering Journal, 452 (2023) 139598.

[23] P.S. Simas, V.S. Antonin, L.S. Parreira, P. Hammer, F.L. Silva, M.S. Kronka, R.B. Valim, M.R.V. Lanza, M.C. Santos, Carbon Modified with Vanadium Nanoparticles for Hydrogen Peroxide Electrogeneration, Electrocatalysis, 8 (2017) 311-320.

[24] L.R. Aveiro, A.G.M. da Silva, V.S. Antonin, E.G. Candido, L.S. Parreira, R.S. Geonmonond, I.C. de Freitas, M.R.V. Lanza, P.H.C. Camargo, M.C. Santos, Carbon-supported MnO2 nanoflowers: Introducing oxygen vacancies for optimized volcano-type electrocatalytic activities towards H2O2 generation, Electrochimica Acta, 268 (2018) 101-110.

[25] J.P.C. Moura, V.S. Antonin, A.B. Trench, M.C. Santos, Hydrogen peroxide electrosynthesis: A comparative study employing Vulcan carbon modification by different MnO2 nanostructures, Electrochimica Acta, 463 (2023) 142852.

[26] V.S. Antonin, L.E.B. Lucchetti, F.M. Souza, V.S. Pinheiro, J.P.C. Moura, A.B. Trench, J.M. de Almeida, P.A.S. Autreto, M.R.V. Lanza, M.C. Santos, Sodium niobate microcubes decorated with ceria nanorods for hydrogen peroxide electrogeneration: An experimental and theoretical study, Journal of Alloys and Compounds, 965 (2023) 171363.

[27] J.F. Carneiro, M.J. Paulo, M. Siaj, A.C. Tavares, M.R.V. Lanza, Nb 2 O 5 nanoparticles supported on reduced graphene oxide sheets as electrocatalyst for the H 2 O 2 electrogeneration, Journal of Catalysis, 332 (2015) 51-61.

[28] A.B. Trench, J.P.C. Moura, V.S. Antonin, T.C. Gentil, M.R.V. Lanza, M.C. Santos, Using a novel gas diffusion electrode based on Vulcan XC-72 carbon modified with

Nb2O5 nanorods for enhancing H2O2 electrogeneration, Journal of Electroanalytical Chemistry, 946 (2023) 117732.
[29] K. Govardhan, A.N. Grace, Metal/Metal Oxide Doped Semiconductor Based Metal Oxide Gas Sensors—A Review, Sensor Letters, 14 (2016) 741-750.
[30] A. Mehtab, J. Ahmed, S.M. Alshehri, Y. Mao, T. Ahmad, Rare earth doped metal oxide nanoparticles for photocatalysis: a perspective, Nanotechnology, 33 (2022).
[31] D. Avnir, Recent Progress in the Study of Molecularly Doped Metals, Adv Mater, 30 (2018) e1706804.
[32] S. Meena, K.S. Anantharaju, Y.S. Vidya, L. Renuka, B. Uma, S.C. Sharma, D. Prasad B, S.S. More, Enhanced sunlight driven photocatalytic activity and electrochemical sensing properties of Ce-doped MnFe2O4 nano magnetic ferrites, Ceramics International, 47 (2021) 14760-14774.
[33] X. Hu, Y. Yu, Z. Sun, Preparation and characterization of cerium-doped multiwalled carbon nanotubes electrode for the electrochemical degradation of low-concentration ceftazidime in aqueous solutions, Electrochimica Acta, 199 (2016) 80-91.
[34] M. Wang, Y. Han, M. Chu, L. Chen, M. Liu, Y. Gu, Enhanced electrochemical performances of cerium-doped Li-Rich Li1.2Ni0.13Co0.13Mn0.54O2 cathode materials, Journal of Alloys and Compounds, 861 (2021) 158000.
[35] J.A. Oliveira, M.O. Reis, M.S. Pires, L.A.M. Ruotolo, T.C. Ramalho, C.R. Oliveira, L.C.T. Lacerda, F.G.E. Nogueira, Zn-doped Nb2O5 photocatalysts driven by visible-light: An experimental and theoretical study, Materials Chemistry and Physics, 228 (2019) 160-167.
[36] E.C. Paz, L.R. Aveiro, V.S. Pinheiro, F.M. Souza, V.B. Lima, F.L. Silva, P. Hammer, M.R.V. Lanza, M.C. Santos, Evaluation of H2O2 electrogeneration and decolorization of Orange II azo dye using tungsten oxide nanoparticle-modified carbon, Applied Catalysis B: Environmental, 232 (2018) 436-445.
[37] O.F. Lopes, E.C. Paris, C. Ribeiro, Synthesis of Nb2O5 nanoparticles through the oxidant peroxide method applied to organic pollutant photodegradation: A mechanistic study, Applied Catalysis B: Environmental, 144 (2014) 800-808.
[38] S.R. Gawali, V.L. Patil, V.G. Deonikar, S.S. Patil, D.R. Patil, P.S. Patil, J. Pant, Ce doped NiO nanoparticles as selective NO 2 gas sensor, Journal of Physics and Chemistry of Solids, 114 (2018) 28-35.
[39] R. Brandiele, M. Zerbetto, M.C. Dalconi, G.A. Rizzi, A.A. Isse, C. Durante, A. Gennaro, Mesoporous Carbon with Different Density of Thiophenic-Like Functional Groups and Their Effect on Oxygen Reduction, ChemSusChem, 12 (2019) 4229-4239.
[40] Y. Zhang, G. Daniel, S. Lanzalaco, A.A. Isse, A. Facchin, A. Wang, E. Brillas, C. Durante, I. Sires, H(2)O(2) production at gas-diffusion cathodes made from agarose-derived carbons with different textural properties for acebutolol degradation in chloride media, Journal of hazardous materials, 423 (2022) 127005.
[41] P. Chamoli, M.K. Das, K.K. Kar, Structural, optical and electronic characteristics of N-doped graphene nanosheets synthesized using urea as reducing agent and nitrogen precursor, Materials Research Express, 4 (2017) 015012.
[42] Y. Zhou, Z. Qiu, M. Lü, A. Zhang, Q. Ma, Preparation and spectroscopic properties of Nb2O5 nanorods, Journal of Luminescence, 128 (2008) 1369-1372.
[43] X. Wei, X. Luo, H. Wang, W. Gu, W. Cai, Y. Lin, C. Zhu, Highly-defective Fe-N-C catalysts towards pH-Universal oxygen reduction reaction, Applied Catalysis B: Environmental, 263 (2020) 118347.
[44] M.H.M.T. Assumpção, R.F.B. De Souza, R.M. Reis, R.S. Rocha, J.R. Steter, P. Hammer, I. Gaubeur, M.L. Calegaro, M.R.V. Lanza, M.C. Santos, Low tungsten content

of nanostructured material supported on carbon for the degradation of phenol, Applied Catalysis B: Environmental, 142-143 (2013) 479-486.

[45] J.F. Carneiro, R.S. Rocha, P. Hammer, R. Bertazzoli, M.R.V. Lanza, Hydrogen peroxide electrogeneration in gas diffusion electrode nanostructured with Ta2O5, Applied Catalysis A: General, 517 (2016) 161-167.

[46] M.H.M.T. Assumpção, A. Moraes, R.F.B. De Souza, M.L. Calegaro, M.R.V. Lanza, E.R. Leite, M.A.L. Cordeiro, P. Hammer, M.C. Santos, Influence of the preparation method and the support on H2O2 electrogeneration using cerium oxide nanoparticles, Electrochimica Acta, 111 (2013) 339-343.

[47] X. Qu, Y. Liu, B. Li, B. Xing, G. Huang, H. Zhao, Z. Jiang, C. Zhang, S.W. Hong, Y. Cao, Nanostructured T-Nb2O5-based composite with reduced graphene oxide for improved performance lithium-ion battery anode, Journal of Materials Science, 55 (2020) 13062-13074.

[48] H. Liu, N. Gao, M. Liao, X. Fang, Hexagonal-like Nb(2)O(5) nanoplates-based photodetectors and photocatalyst with high performances, Scientific reports, 5 (2015) 7716.

[49] S. Zhang, S. Zhang, L. Song, Super-high activity of Bi3+ doped Ag3PO4 and enhanced photocatalytic mechanism, Applied Catalysis B: Environmental, 152-153 (2014) 129-139.

[50] H. Wang, N. Zhang, G. Cheng, H. Guo, Z. Shen, L. Yang, Y. Zhao, A. Alsaedi, T. Hayat, X. Wang, Preparing a photocatalytic Fe doped TiO2/rGO for enhanced bisphenol A and its analogues degradation in water sample, Applied Surface Science, 505 (2020) 144640.

[51] V.S. Pinheiro, E.C. Paz, L.R. Aveiro, L.S. Parreira, F.M. Souza, P.H.C. Camargo, M.C. Santos, Mineralization of paracetamol using a gas diffusion electrode modified with ceria high aspect ratio nanostructures, Electrochimica Acta, 295 (2019) 39-49.

[52] P.J.M. Cordeiro Junior, A.S. Martins, G.B.S. Pereira, F.V. Rocha, M.A.R. Rodrigo, M.R.d.V. Lanza, High-performance gas-diffusion electrodes for H2O2 electrosynthesis, Electrochimica Acta, 430 (2022) 141067.

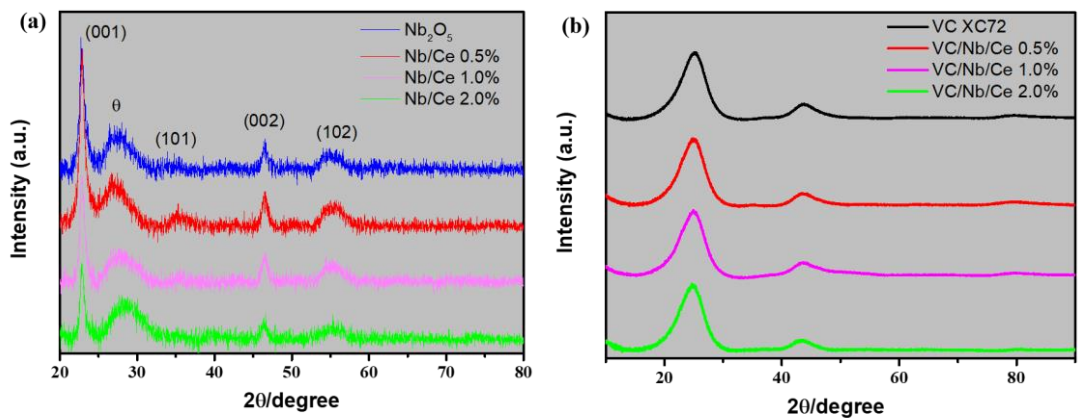

**Fig. 1.** XRD patterns of (a) $Nb_2O_5$ and Nb/Ce with different amounts of Ce and (b) VC XC72 pure and VC/Nb/Ce with different amounts of Ce.

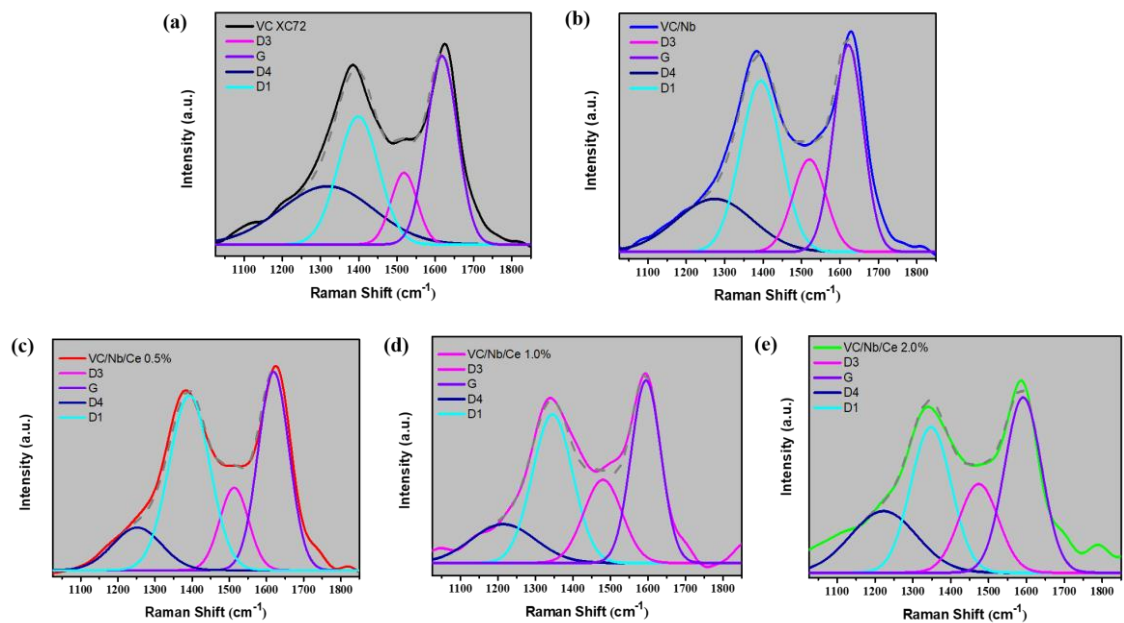

**Fig. 2.** Deconvolved Raman spectra for the electrocatalysts (a) VC XC72, (b) VC/Nb, (c) VC/Nb/Ce 0.5%, (d) VC/Nb/Ce 1.0% and, (e) VC/Nb/Ce 2.0 %.

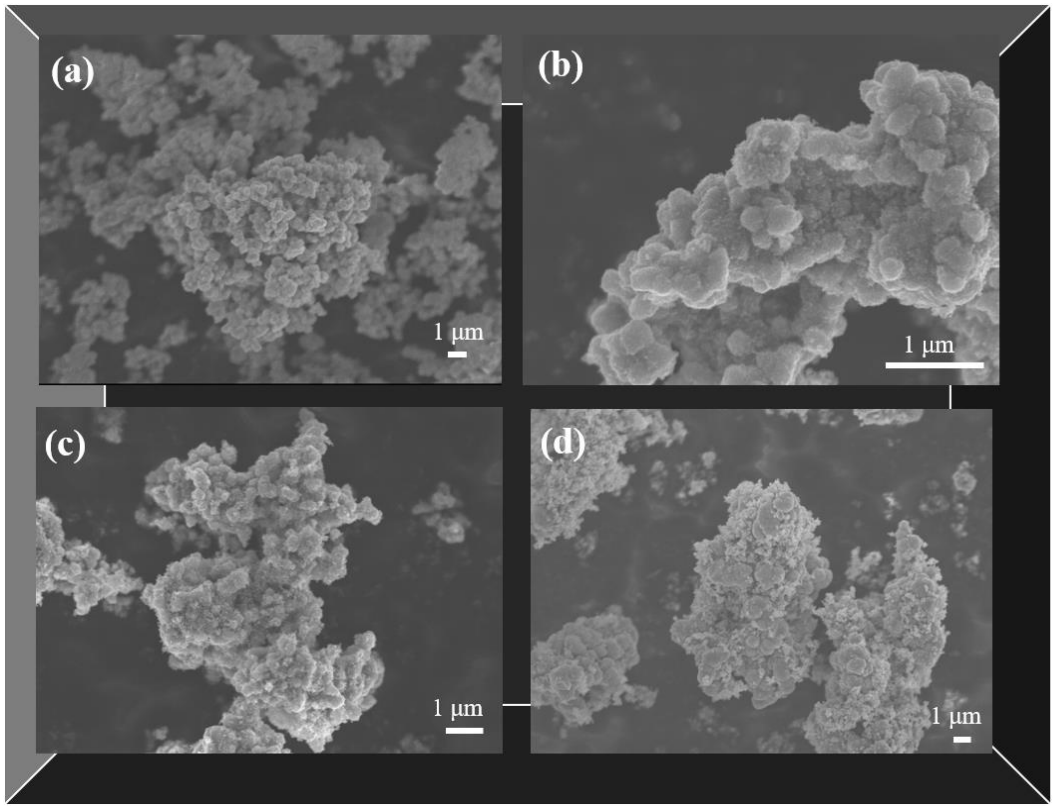

**Fig. 3.** SEM images of (a) Nb$_2$O$_5$, (b) Nb/Ce 0.5%, (c) Nb/Ce 1.0 %, and (d) Nb/Ce 2.0%.

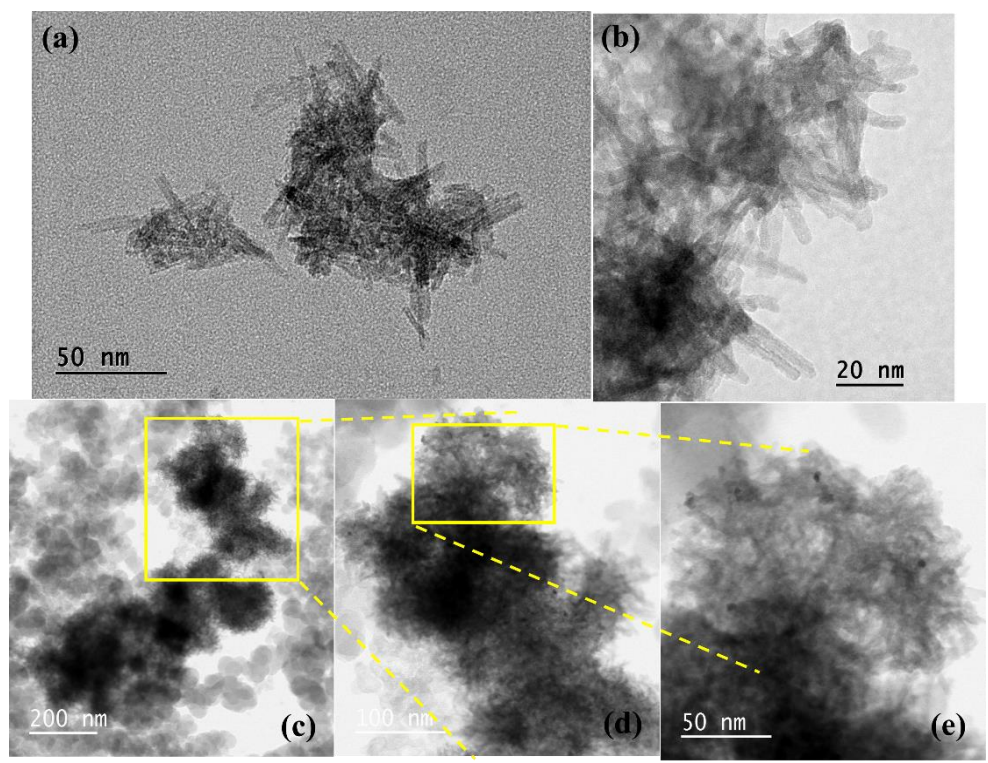

**Fig 4.** Different magnification TEM images of (a-b) Nb$_2$O$_5$ and (c-e) VC/Nb/Ce 2.0%.

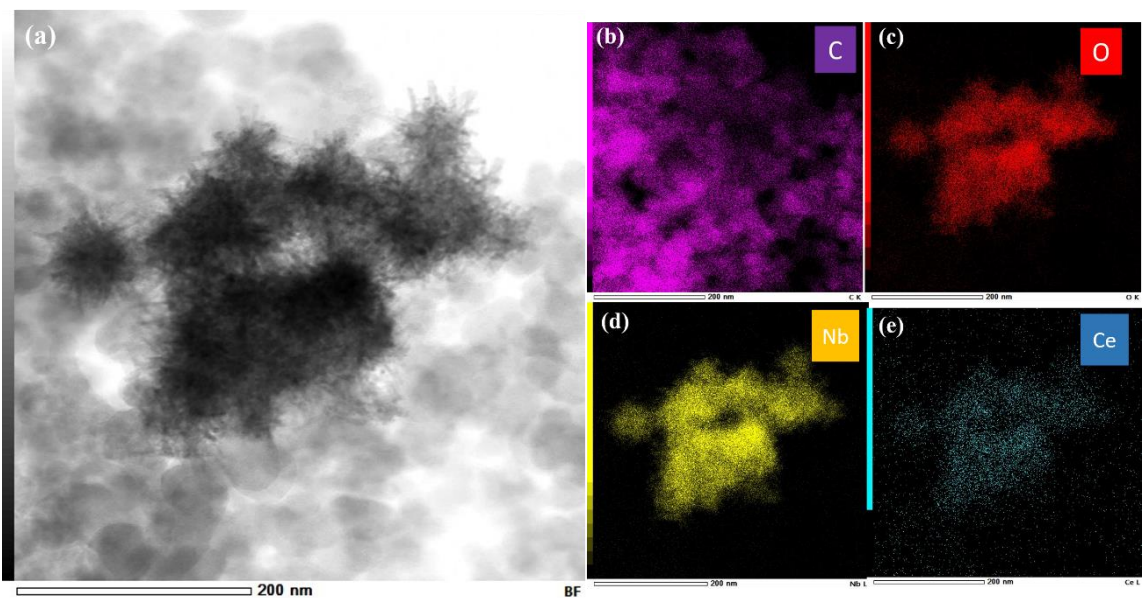

**Fig. 5.** EDS elemental mapping of C, O, Nb, and Ce for the sample VC/Nb/Ce 2.0%.

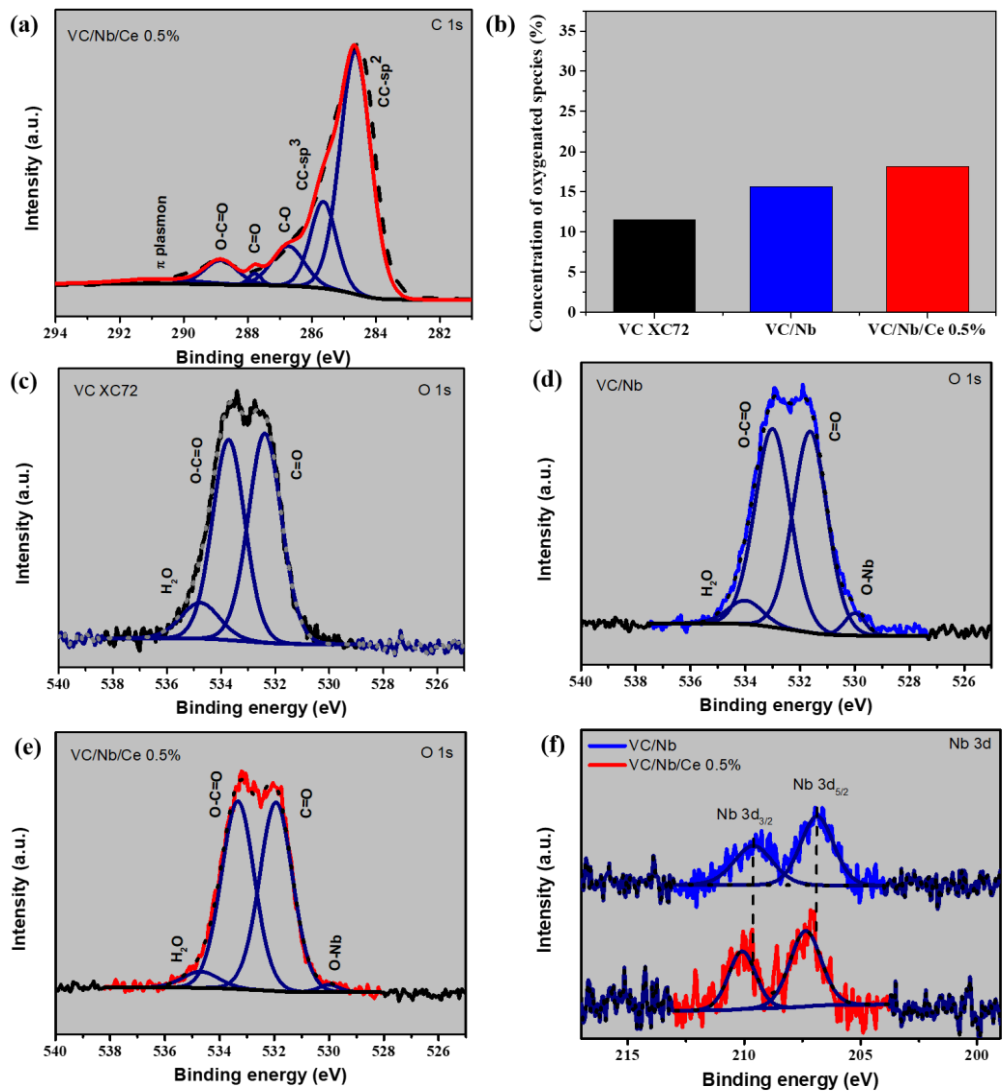

**Fig. 6.** Deconvoluted C 1s XPS spectra of VC/Nb/Ce 0.5% (a), percentage of oxygenated species for the VC XC72 [28], VC/Nb [28] and VC/Nb/Ce 0.5% electrocatalysts (b), deconvoluted O 1s XPS spectra for VC XC72 (c), VC/Nb (d) and VC/Nb/Ce 0.5% (e) and, deconvoluted Nb 3d XPS spectra for VC/Nb and VC/Nb/Ce 0.5% (f).

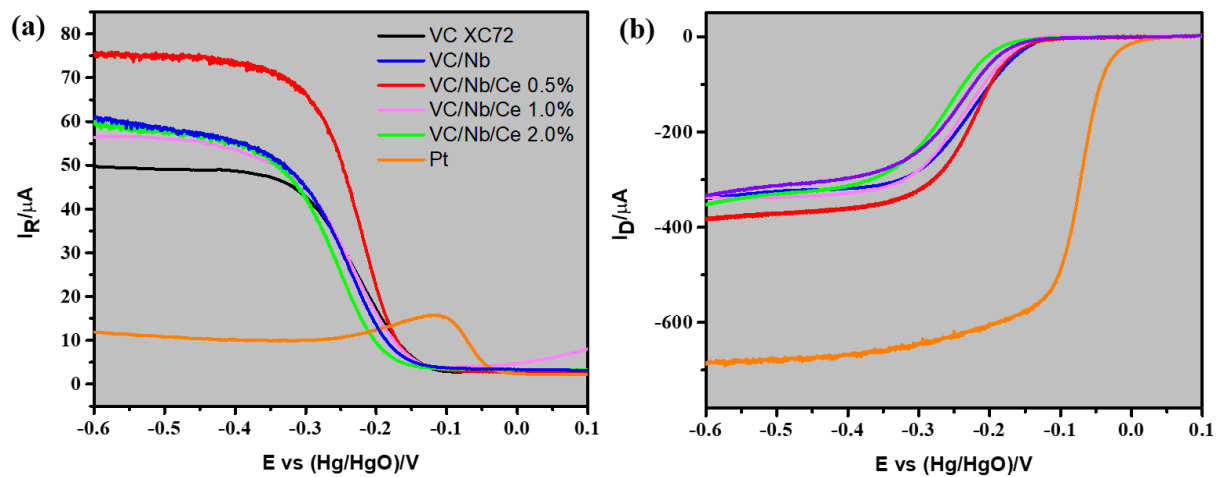

**Fig. 7.** Steady-state polarization curves for ORR electro-catalyzed by electrocatalysts: VC XC72, VC/Nb, Pt, VC/Nb/Ce 0.5%, VC/Nb/Ce 1.0%, and VC/Nb/Ce 2.0% in 1 mol L$^{-1}$ NaOH saturated with O$_2$ at a sweep rate of 5 mV s$^{-1}$ using a RRDE at 1600 rpm: (a) Ring current (Ering = 0.3 V ) and (b) Disc current.

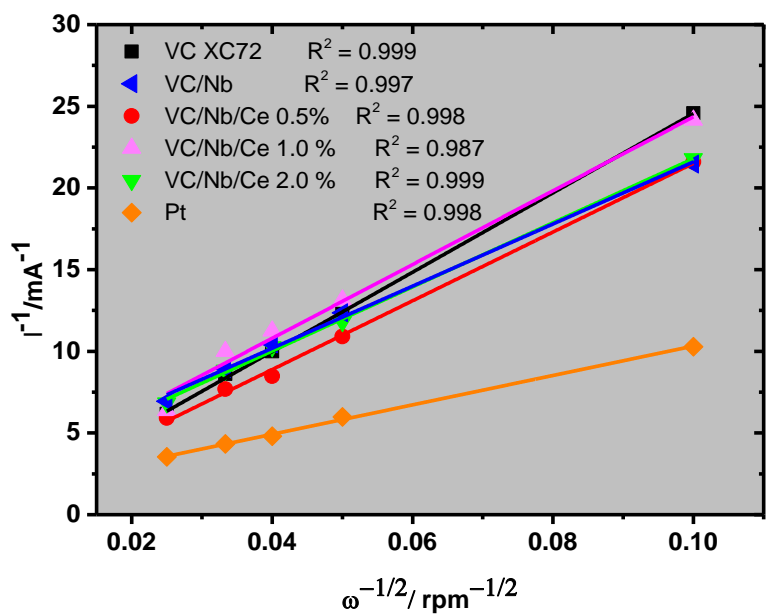

**Fig. 8.** K-L plots for ORR based on the application of VC XC72, VC/Nb, Pt, VC/Nb/Ce 0.5%, VC/Nb/Ce 1.0%, and VC/Nb/Ce 2.0% in 1 mol L$^{-1}$ NaOH saturated with O$_2$.

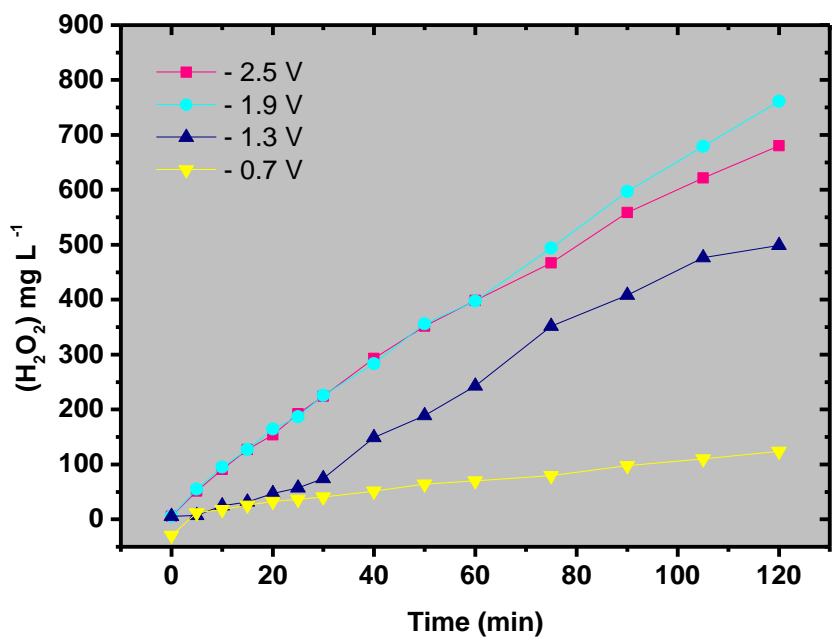

**Fig. 9.** H$_2$O$_2$ concentration obtained from applying VC/Nb/Ce0.5GDE at different potentials in 0.1 mol L$^{-1}$ H$_2$SO$_4$ and 0.1 mol L$^{-1}$ K$_2$SO$_4$ at 20°C after 120 min of electrolysis.